\documentclass[11pt]{article}
\usepackage{hyperref}
\pdfoutput=1

\begin{document}
\title{Secondary jets from high viscosity complex fluid}
\author{K. Bergvall, W. Holm, and G. M{\aa}rtensson \\
\\\vspace{6pt} Micronic Mydata AB, \\ PO Box 3141, SE-183 03 T{\"a}by, Sweden}
\maketitle
\begin{abstract}
The behaviour of a high viscosity complex fluid upon impact with a dry solid surface has been studied. Video-based studies show the appearance of secondary jets for large impact velocities.
\end{abstract}

\section{Introduction}
Jetting of high viscosity, non-Newtonian, particle filled fluids (e.g. solder paste) has been known for some time. Detailed studies of corona splash have been performed for homogeneous Newtonian fluids\footnote{Xu, L. et al. 2005 "Drop Splashing on a Dry Smooth Surface", {\it PRL}, {\bf 94}, pp. 1-4}. Here we study the impact dynamics at high velocities, i.e. \( v > 50 \) m/s . The nozzle is placed over a rotating aluminum wheel at a distance of 0.65 mm and the impact is recorded with a with a CCD camera. A photo diode is used to create a short flash. The fluid that is jetted is a strongly shear thinning, lead free, solder paste with about 45 volume-\% particles with a particle size in the range 15 to 25 $\mu$m. Bursts of 1000 dots were fired at 200 Hz. 

For high impact velocities, secondary jets are emitted after impact. These can be seen as small droplets emitted from the rim of the dot while spreading out on the surface. Sometimes a thin string connected to the emitted droplet can be observed. The emitted droplets typically land somewhere outside the main dot, creating a so called satellite. These observations are important in that they indicate an upper limit on the impact velocity that can be used in a real application.

\end{document}